\documentclass[a4paper]{jpconf}
\usepackage{graphicx}
\usepackage{epsfig}
\usepackage{amsmath}
\usepackage{amssymb}

\newcommand{\GeV}{\mbox{GeV}}

\newcommand{\OO}{\mathcal{O}}

\newcommand{\OP}{\left<\mathcal{O}_P\right>}
\newcommand{\OS}{\left<\mathcal{O}_S\right>}
\newcommand{\RR}{|R'(0)|^2}

\begin{document}
\title{Analytical calculation of heavy quarkonia production processes in computer}

\author{\underline {V.V. Braguta}, A.K. Likhoded, A.V. Luchinsky, S.V. Poslavsky}

\address{IHEP, Pobedy str. 1, Protvino, Moscow region, Russia, 142281}



\begin{abstract}


This report is devoted to the analytical calculation of heavy quarkonia production processes in modern 
experiments such as LHC, B-factories and superB-factories in computer. Theoretical description of heavy 
quarkonia is based on the factorization theorem. This theorem leads to special structure of the production 
amplitudes which can be used to develop computer algorithm which calculates these 
 amplitudes automatically. This report is devoted to the description of this algorithm. As an example of 
its application we present the results of the calculation of double charmonia production in bottomonia decays 
and inclusive the $\chi_{cJ}$ mesons production in pp-collisions.


\end{abstract}

\section{Introduction}

It is well known that the study of heavy quarkonia lead to a great boost in the understanding of QCD \cite{Novikov:1977dq}. 
This is connected to the fact that heavy quarkonia contains all manifestations of QCD. At the same time 
heavy quarkonia are nonrelativistic systems what makes the physics of heavy quarkonia much simpler 
that that for other mesons. For this reason so far the study of heavy quarkonia in different experiments 
is very important and intensive. 

The study of heavy quarkonia production processes is based on the fact that the mass of heavy quark
inside quarkonia is much larger than characteristic QCD energy scale ( $m_Q \gg \Lambda_{QCD}$ ).
This fact leads to the appearance of small parameter -- the relative velocity of quark-antiquark pair 
inside quarkonia $v \ll 1$. For the charmonia mesons this parameter is $v \sim 0.3$, 
for bottomonia mesons $v \sim 0.1$. The appearance of the small parameter separates different 
energy scales involved in quarkonia physics and allows to build a rigorous formalism for studying 
quarkonia properties which is called nonrelativistic QCD (NRQCD) \cite{Bodwin:1994jh}.

\subsection{NRQCD formalism}

NRQCD is based on the separation of different energy scales through the factorization theorem
\begin{equation}
T(e^+ e^- \to \eta_c \gamma)=\sum_n C_n \times \langle \eta_c | \hat O_n  | 0 \rangle,
\label{fact1}
\end{equation}
where $T$ is the amplitude of the process $e^+ e^- \to \eta_c \gamma$\footnote{For definiteness 
 below we consider the process $e^+ e^- \to \eta_c \gamma$}, $C_n$ are the Wilson coefficients 
which parametrize the physics of small distances, $\langle \eta_c | \hat O_n  | 0 \rangle$ are 
different NRQCD operators which contribute to the process. The sum in (\ref{fact1})
is infinite and it runs over all possible operators with correct quantum numbers. In particular, for the production of the 
$\eta_c$ meson the following operators contribute
\begin{equation}
\hat {O} \sim \chi^+ \psi, \chi^+ {\bf D^2} \psi, \chi^+ {\vec H \vec \sigma} \psi, ...
\nonumber
\end{equation}
However, if we restrict the study of the process under consideration by some power of 
relative velocity $O(v^n)$ then only finite number of the NRQCD operators contibute to (\ref{fact1}). 
In particular, at the leading order approximation in relative velocity only the 
$\langle \eta_c | \chi^+ \psi |0 \rangle$ operator contributes.

  If we  omit the operators which assumes that quarkonia consist of more than 
quark-antiquark pair ( for instance, $\chi^+ {\vec H \vec \sigma} \psi$) then 
factorization theorem (\ref{fact1}) can be rewritten in the following form

\begin{equation}
 T(e^+e^- \to \eta_c \gamma) =  
\langle \eta_c | \chi^+ \psi| 0 \rangle \biggl ( C_0 + C_2 \langle q^2 \rangle + C_4 \langle q^4 \rangle + ... \biggr ),
\label{fact2}
\end{equation}
where the matrix elements $\langle q^{n} \rangle$ which control relativistic corrections 
can be written as follows
\begin{equation}
\langle q^{n} \rangle= m_c^n \langle v^n \rangle=\frac {\langle \eta_c | \chi^+ {(- \frac i 2 \bf D)^n} \psi| 0 \rangle} {\langle \eta_c | \chi^+ \psi| 0 \rangle}
\end{equation}
The first few matrix elements $\langle v^n \rangle$ for the $1S, 2S, 1P$ charmonia mesons where calculated 
within QCD sum rules in papers \cite{Braguta:2006wr, Braguta:2007fh, Braguta:2007tq, Braguta:2008qe}.

The Wilson coefficients $C_n$ which parametrize short distance contribution to the amplitude can be expanded 
in the strong coupling constants $\alpha_s$
\begin{equation}
C_n=c_n^{(0)}+c_n^{(1)} \alpha_s + c_n^{(2)} \alpha_s^2+...
\end{equation}
The first aim of the present report is the description of the algorithm of 
analytical calculation of the coefficient $c_n^{(0)}$ for any process. 

\subsection{Light cone expansion formalism}

If a heavy quarkonium production process contains energy scale $E_h$ which is much larger than the masses 
of final mesons, one can apply light cone expansion formalism (LCF) to study such process \cite{Lepage:1980fj, Chernyak:1983ej}. For instance, 
LCF can be applied if the energy of $e^+e^-$ beam is much larger than the masses of charmonia mesons, 
what takes place at B-factories.  The presence of high energy scale $E_h$, which is of order of  
characteristic energy of the hard exclusive process, allows one to apply factorization theorem for the
amplitude of the process (\ref{fact1}). 
As in the case of NRQCD, the coefficients $C_n$ describe partons production at small distances, 
the matrix elements $\langle \eta_c | O_n | 0 \rangle$ describe hadronization of the partons which takes 
place at large distances. The sum is taken over all possible operators $O_n$. For instance, 
the operators 
\begin{equation}
\hat O_n \sim \bar Q \gamma_{\mu} \gamma_5 Q, 
\bar Q \gamma_{\mu} \gamma_5 { { D}_{\mu_1}} { { D} }_{\mu_2} Q, 
\bar Q \sigma_{\mu \nu} \gamma_5 G_{\alpha \beta} Q
\end{equation} 
are few examples of the operator $O_n$. 
Actually, there are infinite 
number of the operators $O_n$ that contribute to the pseudoscalar meson production. 

The cross section of hard exclusive process can be expanded in the inverse powers of the high
energy scale $E_h$
\begin{equation}
\sigma = \frac {\sigma_0} { E_h^n} + \frac {\sigma_1} { E_h^{n+1}} + ... 
\label{exp}
\end{equation}
To determine if some operator contributes to a given term in $1/E_h$ expansion 
one uses the concept of the twist of this operator \cite{Braun:2003rp}. Thus only  the leading twist 
-- the twist-2 operators $\bar Q \hat z \gamma_5 Q, \bar Q \hat z \gamma_5 ({z  { D} })Q, 
\bar Q \hat z \gamma_5 ({z  { D} })^2 Q, ...$\footnote{$z$ here is lightlike fourvector $z^2=0$.}
contribute to the leading term in expansion (\ref{exp}). 
From this one sees that already at the leading order approximation infinite number of operators 
contribute to the cross section. Nevertheless, it is possible to cope with infinite 
number of contributions if one parametrizes all  the twist-2 operators by the moments of some function $\phi(x)$
\begin{equation}
\langle \eta_c(q) | \bar Q \hat z \gamma_5 
({-i  { D} }_{\mu_1})... ({-i  { D} }_{\mu_n}) Q  |0 \rangle \times z^{\mu_1}...z^{\mu_n}  = 
i f_{\eta} (qz)^{n+1} \int_{0}^1  dx \phi(x) (2x-1)^n,
\end{equation}
where $q$ is the momentum of the pseudoscalar meson $\eta_c$, $f_\eta$ is the constant which is defined as
$\langle \eta_c(q) | \bar Q \gamma_{\mu} \gamma_5 Q |0 \rangle = i f_{\eta} q_{\mu}$, $x$ is the 
fraction of momentum of the whole meson $\eta_c$ carried by quark. The function $\phi(x)$ is called the leading 
twist distribution amplitude (DA). One can think of it as about the Fourier transform of the  wave function 
of the meson $\eta_c$ with lightlike distance between quarks.  Using the definition of this function, factorization theorem (\ref{fact1})
can be rewritten as
\begin{equation}
T=\int_0^1 dx H(x) \phi(x),
\label{amp}
\end{equation}
where $H(x)$ is the hard part of the amplitude, which describes small distance effects.
This part of the amplitude can be calculated within perturbative QCD. As it was 
noted, the leading twist DA parametrizes infinite set of the twist-2 operators. 
This part of the amplitude describes hadronization of quark-antiquark pair at large distances
and parametrizes nonperturbative effects in the amplitude. DA of charmonia mesons
were studied in papers \cite{Braguta:2006wr, Braguta:2007fh, Braguta:2007tq, Braguta:2008qe}.

 It should be noted 
that formula (\ref{amp}) resums the contributions of all twist-2 operators. If the meson $\eta_c$ 
is a nonrelativistic meson, formula (\ref{amp}) resums relativistic corrections to the amplitude $T$. 
It should be also noted that if one takes into account the renormalization group running of the 
DA $\phi(x)$, formula (\ref{amp}) resums all leading logarithmic corrections to the amplitude 
of the process under considerations.  

The second aim of the present report is the description of the algorithm of 
analytical calculation of the hard part of the amplitude $H(\xi)$ for any process. 

\subsection{Hadronic production of $P$-wave charmonium states\label{sub:HardChi}} 

Another example is hadronic production of $P$-wave charmonium states at high energies. In NRQCD formalism $\chi_{cJ}$ mesons are considered as nonrelativistic quark-antiquark pairs in color singlet (CS) or color octet (CO) states
\begin{eqnarray}
\left|\chi_{cJ}\right\rangle  & \sim & \RR \left|c\bar{c}\left[^{3}P_{J}^{[1]}\right]\right\rangle +
\OS \left|c\bar{c}\left[^{3}S_{1}^{[8]}\right]g\right\rangle +\OP
\left|c\bar{c}\left[^{3}S_{0}^{[8]}\right]g\right\rangle +\dots,\label{eq:fock}
\end{eqnarray}
The hadronization probability of each component are described by long distance matrix elements $\RR$, $\OS$, $\OP$, that are usually determined from fit of experimental data.

\begin{figure}
\begin{centering}
\includegraphics{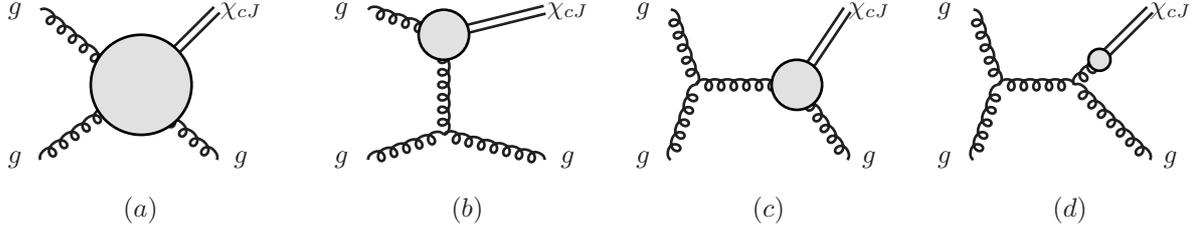}
\par\end{centering}
\caption{Feynman diagrams for $gg\to\chi_{cJ}g$ subprocess\label{diags}}
\end{figure}

In high energy limit gluon-gluon interactions give the main contributions to hadronic production of $\chi_c$-mesons (see diagrams shown in Fig.\ref{diags}). To obtain the analytic expressions for the amplitudes of these processes one can write down the amplitude for $c\bar c$ pair production in the reaction $gg\to c\bar c g$ and project it on the corresponding state. For example, in the case of color-singlet $\chi_{c0}$ meson one have the expression
\begin{eqnarray}
 \mathcal{A}\left(gg\to\chi_{c0} g\right) &\sim& \frac{\partial}{\partial q_\mu} \frac{\partial}{\partial \epsilon_{S\nu}} \mathcal{A}(gg\to c\bar c g) 
\left(\frac{P_\mu P_\nu}{P^2} - g_{\mu\nu}\right),
\end{eqnarray}
where $P_\mu$, $q_\mu$, and $\epsilon_{S\nu}$ are the momentum of final charmonium meson, relative momentum of quarks and spin polarization vector.

\section{Analytical calculation}

The main idea of the algorithm for analytical calculation within NRQCD and LCF is based on the 
factorization theorem (\ref{fact1}). This theorem allows us to make factorization in the 
calculation of the amplitude which consists in the following. The amplitude is devided into 
two part. The first part parametrizes the small distance contribution. The calculation of this 
part is reduced to the calculation of some set of Feynman diagrams. The second part of the amplitude
describes hadronization of quarks. In the calculation it is reduced to the some projection 
operators. In order to get total amplitude these two parts are convoluted by Dirac and Color indices.
To carry out the calculation we use the Mathematica package FeynCalc \cite{feyncalc} for analytical calculation 
in high energy physics.      

%
%
%
%
%

As an example let us consider  analytical calculation of $\langle H(p) \gamma(k) | J_{em}^{\mu} |0 \rangle$ matrix elements,
which can be used to build the whole amplitude of the processes $e^+e^- \to H(p) \gamma(k), H=\eta_c, \chi_{c0}, \chi_{c1}, \chi_{c2}$.
The calculation will be carried out within LCF. Analogously one can repeat the algorithm for 
 NRQCD. 

The program which realizes the algorithm starts from the kinematics 
 \begin{itemize}
   \item There is no transverse motion in the meson H ( $p^{\mu}=\frac {\sqrt s} 2 (1,0,0,1)$ )
   \item The $c$-quark momentum is $p_{Q1} = x_1 p$, the $\bar c$-antiquark momentum $p_{Q2} = x_2 p$ ($x_1+x_2=1$)
   \item The relative momentum of quark-antiquark pair is $q=\xi p,~~ \xi=x_1-x_2$
 \end{itemize}
and the definitions of momenta involved in the process\\ \\
{\bf
p[a\_] = FourVector[p, a];     (*  H - momentum    *)  \\
p1[a\_] = x1*FourVector[p, a];  (*  p1=x1*p  quark momentum    *)  \\
p2[a\_] = x2*FourVector[p, a];  (*  p2=x2*p antiquark momentum  *)  \\
k[a\_] = FourVector[k, a];     (*  photon momentum  *)  \\
A[a\_] = FourVector[A, a];    (* photon polarization *)  \\
\\
ScalarProduct[k, k] = 0; \\
ScalarProduct[p, p] = 0; \\
ScalarProduct[p1, p1] = 0; \\
ScalarProduct[p2, p2] = 0; \\
ScalarProduct[p, k] = s/2; \\
ScalarProduct[p1, k] = x1*s/2; \\
ScalarProduct[p2, k] = x2*s/2; \\
ScalarProduct[A, k] = 0; \\
\\
(* Dirac matrixes *) \\
hp = GS[p]; \\
hp1 = x1*GS[p]; \\
hp2 = x2*GS[p]; \\
hk = GS[k]; \\
hA = GS[A]; \\ }
\\
After the definitions one can calculate the first part of the amplitude -- Feynman diagrams 
of the processes involved \\ \\
{
\bf
T = qc$^2$*e$^2$*ExpandScalarProduct[SpinorUBar[p1].GA[mu].(-hp2 - hk). \\
        hA.SpinorV[p2]/(ScalarProduct[p2 + k, p2 + k]) + \\
        SpinorUBar[p1].hA.(hp1 + hk).GA[mu].SpinorV[p2] \\ /(ScalarProduct[p1 + k, p1 + k])]*
   (SUNTrace[1]/SUNTrace[1]) /. \{SUNN -$>$ 3\}
} \\ \\
Note that thus calculated diagrams  are very simple. If the corresponding 
diagrams are complicated and the number of diagrams is large then it is better to use 
package FeynArt ( which is a part of FeynCalc \cite{feyncalc}), where it is possible to calculate the diagrams automatically. 

The second part of the amplitude which parametrizes hardonization of quarks is represented
by projection operators 
 \begin{itemize}
  \item  the $\eta_c$ meson: $P_{\beta \alpha}  = (\hat p \gamma_5)_{\beta \alpha} \frac {f_{P}}{4}$
  \item  the $\chi_{c0}$ meson: $P_{\beta \alpha}  = (\hat p)_{\beta \alpha} \frac {f_{\chi0}}{4}$
  \item  the $\chi_{c1}$ meson: $P_{\beta \alpha}  = (\hat p \gamma_5)_{\beta \alpha} \frac {f_{\chi1}}{4}$
  \item  the $\chi_{c2}$ meson: $P_{\beta \alpha}  = (\hat p )_{\beta \alpha} \frac {f_{\chi2}}{4}$
 \end{itemize}
It is comfortable to write these  projection operators in terms of the array {\bf Project}
\\ \\
{ \bf
Project = \{fP/4*hp.GA[5], fc0/4*hp, fc1/4*hp.GA[5], fc2/4*hp\}
}\\ \\
Now two parts of the amplitude can be convoluted by Dirac and color indices \\ \\
{\bf
Factor[Tr[T.Project[[1]]]] /. \{x1 + x2 -$>$ 1\} \\
Factor[Tr[T.Project[[2]]]] /. \{x1 + x2 -$>$ 1\} \\
Factor[Tr[T.Project[[3]]]] /. \{x1 + x2 -$>$ 1\} \\
Factor[Tr[T.Project[[4]]]] /. \{x1 + x2 -$>$ 1\} \\
} \\ \\
At the end of the program one gets correct amplitudes of the processes involved. 

 \begin{table}
\begin{center}
\begin{tabular}{|c|c|c|c|c|}
\hline
        & $Br\cdot 10^{-5}$ & $Br\cdot 10^{-5}$  & $Br\cdot 10^{-5}$ & $Br\cdot 10^{-5}$  \\
        & NRQCD \cite{1} & NRQCD \cite{2}  & \cite{Braguta:2009df} & Exp. \cite{Shen:2012ei}  \\
\hline
$\chi_{b0}\to2J/\psi$ & $0.5$ & $1.9$ &  $1.9\pm0.5$ &  $<7.1$  \\
$\chi_{b2}\to2J/\psi$ & $3.4$ & $17.5$ &  $4.8\pm1.2$ &  $<4.5$  \\
\hline
$\chi_{b0}\to J/\psi\ \psi(2S)$& --- & --- &  $2.5\pm 0.7$ &  $<12$  \\
$\chi_{b2}\to J/\psi\ \psi(2S)$& --- & --- &  $5.7\pm 1.8$ &  $<4.9$  \\
\hline
$\chi_{b0}\to 2\psi(2S)$& --- & --- &  $0.8\pm 0.4$ &  $<3.1$  \\
$\chi_{b2}\to 2\psi(2S)$& --- & --- &  $1.7 \pm 0.8$ &  $<1.6$  \\
\hline
\end{tabular}
\end{center}
\caption{The results of the calculation of the bottomonia decays to double charmonia.} 
\label{tab1}
\end{table}.

\section{Applications}

\subsection{Double charmonia production in exclusive bottomonia decays}

Using the algorithm described above in \cite{Braguta:2009df}  all leading 
twist double charmonia production in exclusive bottomonia decays were calculated. The calculation was done
both for NRQCD and LCF. It should be stressed that one program calculated the amplitudes 
and branching ratios for approximately 30 different processes of bottomonia decays. 
We are not going to present all the results obtained in paper \cite{Braguta:2009df} since 
it will take a lot of space. The only result that is presented in this 
report (see Table \ref{tab1}) is the branching ratio for the processes measured at 
Belle collaboration \cite{Shen:2012ei}. It is seen from Table \ref{tab1} that our results (fourth column) 
are in agreement with the results obtained at Belle experiment (fifth column). 
It will be highly desirable to trace the origin of the disagreement in the results of papers 
\cite{Braguta:2009df,1,2}

\subsection{Heavy quarkonia production in hadronic collisions}

\begin{figure}[h!]
\centerline{\includegraphics[width=0.8\textwidth]{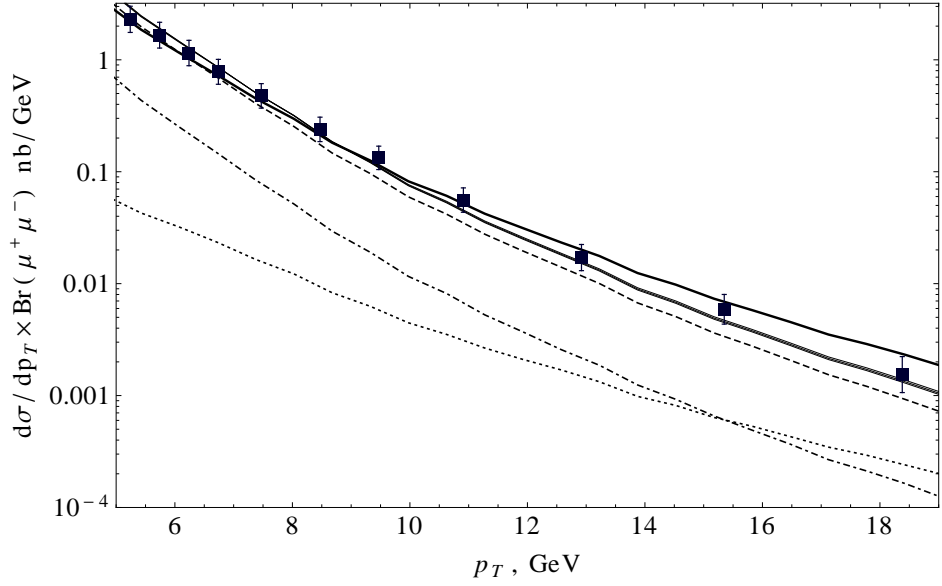}}
\caption{Transverse momentum distribution of $J/\psi$ production in radiative
$\chi_{cJ}$ decays at CDF in comparison with experimental data 
\cite{Abe:1997yz}. Solid lines correspond to the total $pp\to\chi_{cJ}+X\to 
J/\psi+X$ cross section in different choices of $\mu^2$ (upper corresponds 
to $\mu^2 = m_T^2$, and lower three lines to other schemes).
Dashed, dotted and dot-dashed lines correspond to CS, $S$-wave octet and 
$P$-wave octet contributions respectively (all at $\mu^2 = m_T^2$).
\label{fig:sigma}}
\end{figure}

\begin{figure}[h!]
\centerline{\includegraphics[width=0.8\textwidth]{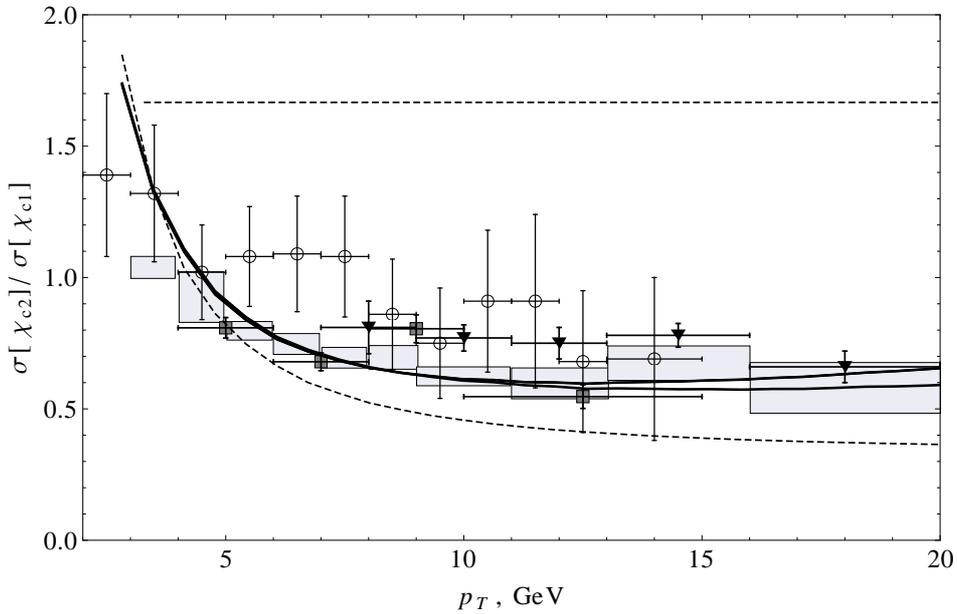}}
\caption{
Theoretical results of the ratio $\chi_{c2}/\chi_{c1}$ in comparison
with experimental data from \cite{Abulencia:2007bra}~($\blacksquare$), 
\cite{LHCb:2012ac}~($\circ$), \cite{CMS21}~($\triangledown$) and 
\cite{Aaij:2013dja} (filled rectangles). Solid lines correspond to 
parameter values presented in Tab. \ref{tab:fit}, while dashed 
lines show predictions with only CS (lower curve) or CO (upper curve) 
contributions taken into account.%
\label{fig:ratio}
}
\end{figure}

Using the formalism described above we have calculated also the cross sections of partonic reactions $gg\to\chi_{cJ}g$
with the help of FeynCalc and Redberry \cite{Bolotin:2013qgr} packages. 
On the basis of these expressions one can calculate the cross sections of hadronic reactions
\begin{eqnarray}
 d\sigma(pp\to\chi_{cJ}+X) &=& \int\limits_0^1 dx_1 dx_2 f_g(x_1) f_g(x_2) d\hat\sigma(gg\to\chi_{cJ}g),
\end{eqnarray}
where $x_{1,2}$ are momentum fractions of initial gluons and $f_g(x)$ is the hadronic structure function. Using the obtained in this way theoretical predictions for the transverse momentum distributions of the color singlet and octet cross sections one can determine from fit to existing experimental data the values of long distance matrix elements $\RR$, $\OS$, $\OP$. Such analysis was performed by our group in papers \cite{Likhoded:2012hw, Likhoded:2013aya}. Presented in these papers numbers are given in Table \ref{tab:fit} and the comparison with experiment is shown in Fig.\ref{fig:sigma} and Fig.\ref{fig:ratio}. One can see, that good agreement is achieved.

\begin{table}
\begin{centering}
\begin{tabular}{|c|c|c|c|}
\hline 
$\mu^{2}$ & $\left|R'(0)\right|^{2},\,\GeV^{5}$ & $\left\langle 
\OO_{S}\right\rangle ,\,\GeV^{3}$ & $\left\langle \OO_{P}\right\rangle 
,\,\GeV^{5}$\tabularnewline
\hline 
\hline 
$M^{2}$ & $0.22$ & $7.6\times10^{-5}$ & $0.031$\tabularnewline
\hline 
$m_{T}^{2}/2$ & $0.34$ & $2.2\times10^{-4}$ & $0.043$\tabularnewline
\hline 
$m_{T}^{2}$ & $0.41$ & $2.7\times10^{-4}$ & $0.053$\tabularnewline
\hline 
$2m_{T}^{2}$ & $0.50$ & $3.3\times10^{-4}$ & $0.065$\tabularnewline
\hline 
\end{tabular}
\par\end{centering}

\caption{CS and CO model parameters for different values of the scale 
$\mu^{2}$\label{tab:fit}.}
\end{table}

\section{Conclusion}

This report is devoted to the analytical calculation of heavy quarkonia production processes in modern 
experiments such as LHC, B-factories and superB-factories in computer. Theoretical description of heavy 
quarkonia is based on the factorization theorem. This theorem leads to special structure of the production 
amplitudes which can be used to develop computer algorithm which calculates these 
 amplitudes automatically. In this report we described this algorithm. As an example of 
its application we presented the results of the calculation of double charmonia production in bottomonia decays 
and inclusive the $\chi_{cJ}$ mesons production in pp-collisions. 

The authors would like to thank the ACAT organization committee for invitation to the conference and hospitality. The work was financially supported by Russian Foundation for Basic Research (grant \#10-00061a), the grant of the president of Russian Federation (\#MK-3513.2012.2), and FRRC grants.


\newpage
\section*{References}
\begin{thebibliography}{9}

\bibitem{Novikov:1977dq} 
  V.~A.~Novikov, L.~B.~Okun, M.~A.~Shifman, A.~I.~Vainshtein, M.~B.~Voloshin and V.~I.~Zakharov,
  Phys.\ Rept.\  {\bf 41}, 1 (1978).

\bibitem{Bodwin:1994jh} 
  G.~T.~Bodwin, E.~Braaten and G.~P.~Lepage,
  Phys.\ Rev.\ D {\bf 51}, 1125 (1995)
  [Erratum-ibid.\ D {\bf 55}, 5853 (1997)]
  [hep-ph/9407339].




\bibitem{Braguta:2006wr} 
  V.~V.~Braguta, A.~K.~Likhoded and A.~V.~Luchinsky,
  Phys.\ Lett.\ B {\bf 646}, 80 (2007)
  [hep-ph/0611021].

\bibitem{Braguta:2007fh} 
  V.~V.~Braguta,
  Phys.\ Rev.\ D {\bf 75}, 094016 (2007)
  [hep-ph/0701234 [HEP-PH]].

\bibitem{Braguta:2007tq} 
  V.~V.~Braguta,
  Phys.\ Rev.\ D {\bf 77}, 034026 (2008)
  [arXiv:0709.3885 [hep-ph]].

\bibitem{Braguta:2008qe} 
  V.~V.~Braguta, A.~K.~Likhoded and A.~V.~Luchinsky,
  Phys.\ Rev.\ D {\bf 79}, 074004 (2009)
  [arXiv:0810.3607 [hep-ph]].


\bibitem{Lepage:1980fj}
  G.~P.~Lepage and S.~J.~Brodsky,
  Phys.\ Rev.\ D {\bf 22}, 2157 (1980).

\bibitem{Chernyak:1983ej}
  V.~L.~Chernyak and A.~R.~Zhitnitsky,
  Phys.\ Rept.\  {\bf 112}, 173 (1984).


\bibitem{Braun:2003rp}
  V.~M.~Braun, G.~P.~Korchemsky and D.~Mueller,
  Prog.\ Part.\ Nucl.\ Phys.\  {\bf 51}, 311 (2003)
  [arXiv:hep-ph/0306057].

\bibitem{feyncalc}
R. Mertig, M. Bohm and A. Denner,
Comp.Phys.Comm. 64, 345 (1991) ; http://www.feyncalc.org/

\bibitem{Braguta:2009df} 
  V.~V.~Braguta, A.~K.~Likhoded and A.~V.~Luchinsky,
  Phys.\ Rev.\ D {\bf 80}, 094008 (2009)
  [Erratum-ibid.\ D {\bf 85}, 119901 (2012)]
  [arXiv:0902.0459 [hep-ph]].

\bibitem{1}
Juan Zhang, Hairong Dong, Feng Feng, Phys.Rev. D84 (2011) 094031 

\bibitem{2}
 Wen-Long Sang, Reyima Rashidin, U-Rae Kim, Jungil Lee, Phys.Rev. D84 (2011) 074026 

\bibitem{Shen:2012ei} 
  C.~P.~Shen {\it et al.}  [Belle Collaboration],
  Phys.\ Rev.\ D {\bf 85}, 071102 (2012)
  [arXiv:1203.0368 [hep-ex]].

\bibitem{Bolotin:2013qgr} 
  D.~A.~Bolotin and S.~V.~Poslavsky,
  [arXiv:1302.1219 [cs.SC]].
  
\bibitem{Likhoded:2012hw} 
  A.~K.~Likhoded, A.~V.~Luchinsky and S.~V.~Poslavsky,
  Phys.\ Rev.\ D {\bf 86}, 074027 (2012)
  [arXiv:1203.4893 [hep-ph]].

\bibitem{Likhoded:2013aya} 
  A.~K.~Likhoded, A.~V.~Luchinsky and S.~V.~Poslavsky,
  arXiv:1305.2389 [hep-ph].

\bibitem{Abe:1997yz} 
  F.~Abe {\it et al.}  [CDF Collaboration],
  Phys.\ Rev.\ Lett.\  {\bf 79}, 578 (1997).

\bibitem{Abulencia:2007bra} 
  A.~Abulencia {\it et al.}  [CDF Collaboration],
  Phys.\ Rev.\ Lett.\  {\bf 98}, 232001 (2007)
  [hep-ex/0703028 [HEP-EX]].

\bibitem{LHCb:2012ac}
 R. Aaij {\it et al.}  [LHCb Collaboration],
  Phys.\ Lett.\ B {\bf 714} (2012) 215  [arXiv:1202.1080 [hep-ex]].


\bibitem{CMS21}
  S.~Chatrchyan {\it et al.}  [CMS Collaboration],
  Eur.\ Phys.\ J.\ C {\bf 72} (2012) 2251  [arXiv:1210.0875 [hep-ex]].

\bibitem{Aaij:2013dja} 
  R.~Aaij {\it et al.}  [LHCb Collaboration],
  arXiv:1307.4285 [hep-ex].
  



\end{thebibliography}
\end{document}